\documentclass[prd,onecolumn,nofootinbib,amsfonts]{revtex4-2}
\usepackage[utf8]{inputenc}
\usepackage[T1]{fontenc}
\usepackage{hyperref}
\usepackage{graphicx,bm,amsmath,color}
\usepackage[dvipsnames]{xcolor}
\usepackage{bbold}
\usepackage{wasysym}
\usepackage{verbatim}
\usepackage{amssymb}

\hypersetup{
colorlinks=true,         
linkcolor=Orange,          
citecolor=OliveGreen,           
urlcolor=Fuchsia   
}

\usepackage{multirow}
\usepackage{tikz}
\usepackage{graphicx}
\usepackage{float}
\usepackage{mathtools}
\usepackage{comment}

\DeclareMathOperator{\arctanh}{arctanh}
\makeatletter

\newcommand{\be}{\begin{equation}}
\newcommand{\ee}{\end{equation}}
\newcommand{\beq}{\begin{eqnarray}}
\newcommand{\eeq}{\end{eqnarray}}
\newcommand{\ba}{\begin{align}}
\newcommand{\ea}{\end{align}}

\newcommand{\mathi}{{\rm i}}

\begin{document}

\title{Non-local quantum field theory from doubly special relativity}
\author{J.J. Relancio}
\affiliation{\small Departamento de Matemáticas y Computación, Universidad de Burgos, Plaza Misael Bañuelos, 09001, Burgos, Spain\\Centro de Astropartículas y F\'{\i}sica de Altas Energ\'{\i}as (CAPA),
Universidad de Zaragoza, C. de Pedro Cerbuna, 12, 50009 Zaragoza, Spain}
\email{jjrelancio@ubu.es}
\author{L. Santamaría-Sanz}
\affiliation{\small {Departamento de Matemáticas y Computación}, Universidad de Burgos, Plaza Misael Bañuelos, 09001 Burgos, Spain}
\email{lssanz@ubu.es}

\begin{abstract}%

Doubly special relativity is usually regarded as a low-energy limit of a quantum gravity theory with testable predictions. On the other hand, non-local quantum field theories have been presented as a solution to the inconsistencies arising when quantizing gravity. Here, we present a new formulation of quantum field theories in doubly special relativity with non-local behavior.  Our construction restricts the models to those showing linear Lorentz invariance. We derive the deformed Klein--Gordon, Dirac, and electromagnetic Lagrangians, as well as the deformed Maxwell equations. We also discuss the electric potential of a point charge. Finally, we analyze the connection between the nonlocality of field theories and doubly special relativity.
\end{abstract}

\maketitle

\section{Introduction} A formulation of a quantum gravity (QG) theory, unifying quantum field theory (QFT) and general relativity (GR), {has been sought over the last decades}. Some of these formulations are loop quantum gravity~\cite{Dupuis:2012yw}, causal set theory~\cite{Wallden:2013kka}, and string theory~\cite{Mukhi:2011zz}. The last two theories lead to non-local effects, showing a completely different behavior with respect to the classical notion of spacetime. 

This non-locality was implemented in quantum field theories by including actions with infinite derivatives~\cite{Efimov:1967pjn,Krasnikov:1987yj,Tomboulis:1997gg}, for both causal set~\cite{Aslanbeigi:2014zva} and string~\cite{Buoninfante:2018mre} theories. These non-local theories show some particular and interesting features. In contrast to other approaches with more 
than two (but with a finite number of) derivatives~\cite{Sotiriou:2008rp}, infinite derivative theories~\cite{Buoninfante:2018mre} are ghost free theories~\cite{Ostrogradsky:1850fid}. This allows us to apply infinite derivatives to  quantum fields without changing the content of fields.  The infinite derivatives are functions of the d'Alembertian operator so linear Lorentz covariance is preserved. These functions of the  d'Alembertian depend on the considered theory, being in string theory an  exponential~\cite{Biswas:2014tua}, and in causal set theory a non-analytical expression~\cite{Aslanbeigi:2014zva}. In some of these theories, ghosts  disappear under certain restrictive conditions, such as the introduction of transcendental (i.e. non-polynomial) entire functions of the covariant D'Alembertian operator in the action~\cite{Krasnikov:1987yj,Tomboulis:1997gg}. Furthermore, in~\cite{Buoninfante:2018mre} the absence of ghosts is guaranteed by the lack of new dynamical degrees of freedom left in the propagator, apart from the original transverse and traceless graviton. Moreover, the notion of causality in nonlocal quantum theories is modified for small scales~\cite{Buoninfante:2018mre,Boos:2018kir}. 
Another feature of these non-local field theories that arises when they are used for describing gravity (the so-called infinite derivative gravity (IDG) theories~\cite{Modesto:2017sdr}) is a non-divergent gravitational potential at the origin~\cite{Boos:2018bxf, Buoninfante:2018stt}.  
 
Due to the difficulties in discussing phenomenological implications of the aforementioned QG theories, ``top-down'' approaches have been considered for describing QG effects at low-energy so that they could be tested. Some of these theories modify the kinematics of special relativity (SR) by introducing a high-energy scale $\Lambda$. When doing so, there are two different possibilities regarding Lorentz invariance. One possibility is to consider that Lorentz invariance is broken for high-energies~\cite{Kostelecky:2008ts}, {losing} the relativity principle that characterizes SR. On the other hand, it is possible to deform the kinematics of SR while keeping an observer invariance of the theory. This is realized in doubly/deformed special relativity (DSR) scenarios~\cite{AmelinoCamelia:2008qg}, which present different phenomenological predictions with respect to the other {aforementioned scheme}~\cite{Addazi:2021xuf}. In particular, in DSR theories the three main pillars of the kinematics of SR, viz. the dispersion relation, Lorentz transformations, and energy-momentum conservation laws, are deformed accordingly to a relativity principle. Therefore, new terms proportional to momenta and the inverse of the high-energy scale are added to the standard SR terms. We can illustrate this with simple examples at lowest order in $\Lambda$: the quadratic dispersion relation of SR could be deformed in this scenario as
\begin{equation}
   C(p)=p^2+ (p^2)^2/\Lambda^2+\cdots\,,
\end{equation}
and the deformed conservation law $\oplus$ {(which we will call deformed composition law)} as 
 \begin{equation}
    (p\oplus q)_\mu=p_\mu + q_\mu (1+p^2/\Lambda^2)+\cdots\,,
\end{equation}
being $p^2=p_\mu \eta^{\mu\nu}p_\nu$, with $\eta=\mathrm{diag}(1,-1,-1,-1)$. {While the Casimir can be the same as in SR \cite{Borowiec2010} (and then the theory preserve linear Lorentz invariance), the deformed composition law for the momenta will never be the sum of individual momenta. Then, the main ingredient of DSR theories is a nontrivial composition law.}

The most studied models of DSR are  $\kappa$-Poincaré~\cite{Lukierski:1991pn} and Snyder kinematics~\cite{Battisti:2010sr}. They differ in the fact that the latter presents a linear Lorentz invariance for a system of multiple particles, which is not the case of the former. Both kinematics can be obtained from an algebraic point of view, but also from geometrical arguments of a maximally symmetric momentum space~\cite{Carmona:2019fwf}, where the deformed dispersion relation is understood as the squared distance in momentum space, and the deformed Lorentz transformations and conservation laws as isometries of such space. A particular possibility is to consider an anti-de Sitter (AdS) momentum space,\footnote{Note that from now on when we talk about AdS and de Sitter (dS) spaces we mean momentum spaces. } whose symmetries have in fact been shown to be present in a QG theory in 2+1 dimensions~\cite{Matschull:1997du,Freidel:2003sp}. The $\kappa$-Poincaré  kinematics cannot be realized for this space, leaving only space to Snyder models,  as discussed in~\cite{Carmona:2019fwf}. 

The interpretation of a deformed kinematics from a curved momentum space proposed in~\cite{Carmona:2019fwf} was used as a guidance in~\cite{Franchino-Vinas:2022fkh} {for constructing} a QFT in momentum space, obtaining some results compatible with previous {algebra-based works}~\cite{Nowicki:1992if}. This fact opened a new way of considering a QFT in DSR by geometrical arguments, the counterpart version of the most studied algebraic approach (see~\cite{Arzano:2020jro} and references therein).  

The aim of this work is to formulate a DSR QFT in position space,  following the {geometrical-based version} in momentum space carried out in~\cite{Franchino-Vinas:2022fkh}. Due to the infinite number of derivatives involved, our proposal leads to a non-local QFT that can be used to effectively describe simultaneously the QFTs derived from string theory, causal set theory, and, for the first time, DSR (see the discussion of~\cite{Relancio:2024loc}). In the DSR setting, we describe how to modify the usual QFT's building blocks in order to take into account the deformed symmetries of DSR. We also explain how the Klein--Gordon, Dirac, and electromagnetic (EM) Lagrangians are modified. For the latter, and considering the particular metric found to be privileged from a geometrical approach to DSR~\cite{Relancio:2020rys}, we find that the electric potential of a point charge does not diverge at the origin for AdS space, while it does for the dS case.

\section{Revisiting DSR QFT in momentum space}
\label{sec:revision}
In~\cite{Relancio:2020zok} it was shown that the   metric and the dispersion relation, {which is interpreted as the squared distance in momentum space}, are related in a simple way by
  \begin{equation}
f^\mu(p) g_{\mu \nu }(p)f^\nu(p)=C(p)\,,
\label{eq:casimir_metric}
  \end{equation}
where 
\begin{equation}
f^\mu  (p)=\frac{1}{2} \frac{\partial C(p)}{\partial p_\mu}\,.
   \label{eq:f_definition}
\end{equation}
In this way, the dispersion relation in the theory can be written as $C(p)-m^2=0$. For SR, the previous equation reduces to the usual quadratic relation in momenta.

 On the one hand, we can describe the Klein-Gordon equation in the momentum space in DSR as~\cite{Franchino-Vinas:2022fkh}:
\begin{equation}
\left(  f^\mu(p) g_{\mu \nu} (p)  f^\nu (p)  -m^2\right) \tilde\phi(p)=0\,,
   \label{eq:KG_DSR}
\end{equation}
being $\tilde\phi(p)$ the Fourier transform of the scalar field $\phi(x)$, i.e.
\begin{equation}
   \phi(x)=\frac{1}{(2\pi)^{3}}\int  {\rm d}^4 p\, e^{\mathi x^\lambda p_\lambda} 
   \tilde \phi(p)\, \delta(C(p)-m^2)
   \,.
   \label{eq:KG_field}
\end{equation}
{This is the simplest Fourier transform that one can consider, which is also the one used in nonlocal QFT~\cite{Efimov:1967pjn,Buoninfante:2018mre}. This differs from the proposal of~\cite{Arzano:2020jro}, where the Fourier transform includes the product of exponentials involving noncommutative spacetime coordinates.  Here we will consider a commutative spacetime with a curved momentum space, in such a way that the usual Poisson structure is maintained. This is the most natural conclusion based on geometric arguments in cotangent bundle geometries in the context of DSR (see~\cite{Relancio:2022mia} and references there in). Following this way of thinking, one could add a geometric (momentum dependent) measure in the Fourier transform, as done in~\cite{Amelino-Camelia:1999jfz}. However, if this deformed measure is considered, an overall momentum dependent factor will appear, which will not change any of the results of this work. }

The corresponding action in the Klein-Gordon theory is  given by
\begin{align}\label{eq:KG_action_p}
S_{\rm KG}\,:&=\,\int {\rm d}^4p\, \sqrt{-g(p)} \, \tilde\phi^*(p) \left(  C(p)  -m^2  \right)   \tilde\phi(p)
\,.
   \end{align}
On the other hand, the Dirac equation in DSR 
  \begin{eqnarray}
\left( \gamma^\mu \eta_{\mu\rho} e^\rho{}_\nu({p}) f^\nu({p})-m\right)\tilde\psi(p)=0\,,
 \label{eq:Dirac_DSR}
\end{eqnarray} and its corresponding action 
\begin{align}
\!\! \!\!  S_{\rm D}\!:=\!\int \!\!  {\rm d}^4p\, \sqrt{-g(p)}  \tilde \psi(-p)\left(  \gamma^\mu \eta_{\mu\rho} e^\rho{}_\nu({p}) f^\nu({p})-m\right) \tilde\psi(p)\,,
\label{eq:Dirac_action_p}
\end{align}
were also obtained in \cite{Franchino-Vinas:2022fkh}. 
Here, $\gamma^\mu$ are the usual Dirac matrices in SR, and $ e^\mu{}_\nu(p)$ the tetrad in momentum space which leads to the metric through
  \begin{equation}
g_{\mu\nu}(p) =e^\rho{}_\mu (p)\eta_{\rho \sigma}e^\sigma{}_\nu (p)\,.
\end{equation}
The choice of the tetrad is related to the composition of momenta in the following way \cite{Franchino-Vinas:2022fkh}
\be
{e}^\mu{}_\nu (p):= \left. \frac{\partial \left(p \oplus q\right)_\nu}{\partial q_\mu}\right|_{q\to 0}\,.
\label{eq:tetrad2}
\ee  
{This relationship can be easily obtained from the fact that the composition law is an isometry of the momentum metric, so~\cite{Carmona:2019fwf}
\begin{equation}
g_{\mu\nu}\left(p\oplus q\right) \,=\,\frac{\partial \left(p\oplus q\right)_\mu}{\partial q_\rho} g_{\rho\sigma}(q)\frac{\partial \left(p\oplus q\right)_\nu}{\partial q_\sigma}
\label{eq:composition_isometry}
\end{equation}
holds. Then, by taking the limit $q\to 0$ one finds Eq.~\eqref{eq:tetrad2}. This relationship between the tetrad and the composition law was  crucial when reproducing the Dirac equation obtained from the Hopf algebra scheme \cite{Nowicki:1992if} from a field theory in momentum space \cite{Franchino-Vinas:2022fkh}. Then, we will use it in the following, associating different tetrads to different kinematics (since they have different composition laws although the same metric).}

Either in Klein-Gordon equation{,} as well as in Dirac one, the term $ \sqrt{-g(p)}$ was introduced in order to have invariance under momentum diffeomorphisms, i.e., change of basis of the kinematics. Moreover, it is important to remark that with this geometrical construction one obtains the same results derived previously in the literature from algebraic considerations inside the Hopf algebra scheme~\cite{Nowicki:1992if}. 

\section{Towards a DSR QFT in position space}
\label{sec:position}
Only the momentum space has been considered so far. The aim of this paper is  to generalize the usual axioms concerning derivatives in position space in general QFT to  DSR. A simple way {to} generalize the Klein--Gordon action is to propose the following {one}
\begin{eqnarray}
&&\!\!\!\!\!\!S=\!\!\int {\rm d}^4 x \frac{1}{2} \left\lbrace  \phi(x)\left( C(-i\partial_x)-m^2 \right) \phi(x) \right\rbrace\,,
\label{eq:KG_action_c}
   \end{eqnarray}
as usually done in non-local QFT~\cite{Buoninfante:2018mre}. Note that in \eqref{eq:KG_action_c}, the replacement $p \to -i\partial_x$ is done in the argument of the Casimir operator, so~\eqref{eq:KG_action_p} is the analogous action in momentum space {(we will momentarily ignore the square root of the determinant of the metric.)}.  

Now we can wonder about how to consider a generalization of the usual Klein--Gordon equation (and then of the Dirac one) by writing an action in which the derivatives apply to both fields, and not only to the second of them. This can be done by using Eq.~\eqref{eq:casimir_metric}, so that one possible answer is to consider
\begin{eqnarray}
S=\!\!\int {\rm d}^4 x \frac{1}{2} &&\left\lbrace \,  { f^\mu (i\partial_x) }\phi(x) g_{\mu \nu} (-i\partial_x)  f^\nu (-i\partial_x) \phi(x)\right. \nonumber\\&&\,\,\, 
\left.-m^2 \phi^2(x) \right\rbrace\,.
\label{eq:KG_action_wrong}
   \end{eqnarray}
Some problems may arise within this approach. In particular, the metric is promoted to a differential operator, making the action of the function of derivatives applied to the first field (i.e. {$f^\mu (i\partial_x) \phi(x)$}) different from that applied to the second (viz. $g_{\mu \nu} (-i\partial_x)  f^\nu (-i\partial_x) \phi(x)$). This makes that whenever one wants to lower or upper indexes, one must use a differential operator (the metric $g_{\mu\nu} (-i\partial_x)$), which can be rather difficult and confusing. A simple way to avoid the previous disadvantages is by considering the following action
\begin{eqnarray}
&&\! S=\!\!\int \!  \frac{{\rm d}^4 x}{2} \left\lbrace  -\ell^\mu(i\partial_x)\,  \phi(x) \eta_{\mu \nu} \ell^\nu(-i\partial_x)\,  \phi(x)  -m^2 \phi^2(x) \right\rbrace,\nonumber\\
&& \textrm{where} \quad \ell^\mu (-i\partial_x)= {e^\mu}_\nu (p) f^\nu (p)\left.\right|_{p \to -i\partial_x}\,.
\label{eq:KG_action}
   \end{eqnarray}
{Note that the action obtained from Eq.~\eqref{eq:KG_action_wrong} should thus be
\begin{eqnarray}
&&\! S=\!\!\int \!  \frac{{\rm d}^4 x}{2} \left\lbrace  {\ell^\mu(i\partial_x)}\,  \phi(x) \eta_{\mu \nu} \ell^\nu(-i\partial_x)\,  \phi(x)  -m^2 \phi^2(x) \right\rbrace\,.
   \end{eqnarray}
However, in this action the derivative functions do not act in a symmetric way under the two fields (this would imply that gauge invariance will be lost when considering the electromagnetic action). Then, we will consider the action of Eq.~\eqref{eq:KG_action}.
Therefore, our proposed construction consists of} replacing the usual derivatives in QFT by a function of them that takes into account the curvature in momentum space, so the deformed dispersion relation is satisfied. In this way, one can interpret the proposed setup as a way to generalize QFT, so DSR deformed symmetries are taken into account. 

In addition, our construction implies that the Minkowski metric is at the base of the proposal.  Notice that the square root of the determinant of the metric was included in~\cite{Franchino-Vinas:2022fkh} in order for the model to present invariance under changes of momentum coordinates. But since our proposal selects some metrics (those with linear Lorentz invariance, as we will discuss), we are losing this property, so the idea of including such a term should be forgotten. In addition, the same linear Lorentz invariance that characterizes the usual QFT should also be present in our scheme, since we are considering the same Minkowski metric. This restricts the possible bases of kinematics, {leaving only space for} those possessing such an invariance. For them, the dispersion relation is a quadratic expression on the momentum, $C=C(p^2)$, and therefore we can write\footnote{In the following, for alleviating the notation, when functions are expressed in terms of $p^2$ we mean $p^2/\Lambda^2$, for dimensional reasons.}
\begin{equation}
f^\mu  (p)=\frac{1}{2} \frac{\partial p^2}{\partial p_\mu}\frac{\partial C(p)}{\partial p^2}=p^\mu \frac{\partial C(p)}{\partial p^2} \coloneqq p^\mu h(p^2)\,.
   \label{eq:f_cov}
\end{equation}

Notice that having a relativity principle implies that Lorentz transformations and the composition law must be isometries of the momentum metric. From a geometrical point of view, linear Lorentz transformations are isometries whenever the metric is of the form
\begin{equation}
g_{\mu\nu}(p)=\eta_{\mu\nu} \, \Theta_1 (p^2)+\frac{p_\mu p_\nu}{\Lambda^2}\,  \Theta_2(p^2)\,.
\label{eq:lorentz_metric}
\end{equation}
where $\Lambda$ is the high{-}energy scale. This kind of metrics was also found to be privileged from a geometrical point of view when a curvature in both momentum space and spacetime is considered~\cite{Relancio:2020rys,Pfeifer:2021tas}. There are different (indeed infinite) choices of momentum metrics that possess linear Lorentz transformations as isometries. In particular, we will focus on a maximally symmetric momentum metric which is conformally Minkowski, 
\begin{equation}
g_{\mu\nu}(p)=\eta_{\mu \nu}\left(1\pm\frac{p^2}{4\Lambda^2}\right)^2\,,
\label{eq:conformal_metric}
\end{equation}
because it is particularly favored from {a geometrical interpretation of DSR due to its conformal form}~\cite{Relancio:2020rys,Chirco:2022jvx}.\footnote{{On the one hand, this metric corresponds to a maximally symmetric space, from which one can construct a deformed relativistic kinematics, as discussed above.  On the other hand, due to its conformal form the same Einstein equations of GR are obtained, even if the metric is momentum dependent. This implies that there is no need of extra (momentum dependent) source terms, very difficult to be interpreted form a physical point of view~\cite{Relancio:2020rys}. Moreover, the Einstein tensor is conserved under the covariant derivative of this metric~\cite{Chirco:2022jvx}. These last two conditions are not necessarily true for other metrics. Then, due to all its properties, we will use this metric in the following. }} Note that the plus sign stands for AdS and the minus for dS. Now, we will generalize the Klein--Gordon, Dirac, and electromagnetic Lagrangians within our approach.  

{Before going on, we would like to comment the differences between our approach and other works of the literature. In~\cite{Battisti:2010sr,Arzano:2020jro}, the authors present a field theory in which a deformed composition law (given by the start product) always appears, even in the free part of the theory. In our work, the free part is only modified by the action of the tilde partial operator. Then, our proposal is to consider that the free part involves infinite derivatives ({thus} capturing the physical implications of a curved momentum space) but does not involve the composition law, which should appear when interactions come into play. Due to this difference between our approach and those of the literature, the free part of the theory (and also the interaction one) should be different. We hope to explore the differences and similarities of both approaches in future works. }

\section{Klein-Gordon equation. }
\label{sec:kg} Applying the variational principle over the  action~\eqref{eq:KG_action} yields the equation of motion (e.o.m.):
\begin{eqnarray}\label{eq20bis}
\left(    \ell^\mu(i\partial_x)\ell_\mu(-i\partial_x) 
 +m^2 \right)\phi(x)=0.
\end{eqnarray}

{The aforementioned dispersion relation $C(p)-m^2=0$  is satisfied only if we impose $\ell^\mu(-i\partial_x)=-\ell^\mu(i\partial_x)$}. From Eq.~\eqref{eq:f_cov} it is easy to see that $f^\mu(-p)=-f^\mu(p)$, so  $e^\mu{}_\nu(-p)=e^\mu{}_\nu(p)$ must hold. It means that the momentum space tetrad must depend only on quadratic terms on $p$. {Hence, the composition law cannot depend on a fixed timelike vector, due to the relation in Eq.~\eqref{eq:tetrad2}. Consequently, $\kappa$-Poincaré kinematics are not allowed}\footnote{The first order term involving the high-energy scale in the composition law of any basis of $\kappa$-Poincaré, which depends on a fixed vector $n^\mu$, is of the form $p\cdot n =p_\mu n^\mu$. Consequently, the associated tetrad depends linearly on $p_\mu n^\mu$, where $n^\mu$ is usually chosen as a timelike vector in order to preserve isotropy in the model. Since the tetrad in our approach must {satisfy} $e^\mu{}_\nu(-p)=e^\mu{}_\nu(p)$, it could depend on $(p\cdot n)^2$, but not in linear terms as the ones appearing in $\kappa$-Poincaré.} in our scheme, leaving only space for Snyder models. {This implies that we can replace $\ell^\mu(-i\partial_x) \to - i\partial^\mu  \Omega(-\Box)$ in the previous equations \eqref{eq:KG_action} and \eqref{eq20bis},} being $\Box=\partial^\mu\partial_\mu$,  and $\Omega$ a function depending on the momentum space tetrad, as well as on the deformed dispersion relation.  Consequently, our construction relays basically on the replacement $\partial^\mu \to \tilde \partial^\mu = \partial^\mu\Omega(-\Box)$ in the standard relativistic  QFT. Thus, the  Klein--Gordon action for real fields in our scheme is 
\begin{equation}
S=\int {\rm d}^4 x\,\,  \frac{1}{2}\left\lbrace  \tilde \partial^\mu \phi(x) \tilde \partial_\mu \phi(x)  -m^2 \phi^2(x) \right\rbrace\,.
\label{eq:KG_action3}
   \end{equation}
   From here, the energy-momentum tensor can be obtained following~\cite{Birrell:1982ix} and takes the form
\begin{equation}
    T_{\mu\nu}=\tilde{\partial}_\mu \phi \tilde{\partial}_\nu  \phi- \frac{1}{2}\eta_{\mu \nu} \left(\tilde{\partial}^\rho \phi \tilde{\partial}_\rho \phi -m^2 \phi^2 \right) \,.
    \label{eq:energy_momentum_tensor}
\end{equation}
Notice that {$\partial^\mu T_{\mu\nu}=0$}, and  $T_{\mu\nu}$ is a conserved (symmetric) energy-momentum tensor.

\section{Dirac equation}
In a similar way, the action for Dirac fields is given by
\begin{align}
 S&=\int {\rm d}^4 x\,   \bar\psi(x)\left(i \gamma^\mu\, \tilde\partial_\mu -m\right)\psi(x)\,.
 \label{eq:Dirac_action}
\end{align}
Computing variations, and using $\bar\psi(x)=\psi^{\dagger}(x)\gamma^0$, one obtains the following e.o.m.:
\begin{align}
\left(i \gamma^\mu  \tilde \partial_\mu -m\right)=0{\,.}
\end{align}
Moreover, the energy-momentum tensor can be written as:
\begin{equation}
 \!\!\!   T_{\mu\nu}=i \bar\psi(x)\gamma_\mu\tilde \partial_\nu \psi(x)-\eta_{\mu \nu}\bar\psi(x)\left(i\gamma^\rho \tilde \partial_\rho -m\right) \psi(x)\,.
    \label{eq:energy_momentum_dirac}
\end{equation}

\section{Electromagnetic Lagrangian}
\label{sec:em}
The deformed Maxwell tensor in our approach is given by
\begin{equation}
\tilde{F}_{\mu\nu}=\tilde \partial_{\mu}  A_{\nu}-\tilde \partial_{\nu} A_{\mu} = \Omega(-\Box) F_{\mu\nu}\,,
   \label{eq:Maxell_tensor}
\end{equation}
being $F_{\mu\nu}$ the {standard} electromagnetic tensor.
The action when adding a minimal coupling to matter of the EM field is
\begin{align}
 S_{EM}&=-\int  {\rm d}^4 x \left(\frac{1}{4}\tilde{F}_{\mu\nu}\tilde{F}^{\mu\nu}+j^\mu A_\mu\right)\,.
 \label{eq:action_EM2}
\end{align}
We see that gauge invariance {is preserved} for the transformation
\begin{equation}
    A^\prime_\mu=A_\mu+\tilde \partial_\mu  \Theta\,.
\end{equation}
Defining the electric and magnetic fields as usual: 
\begin{equation}
E_i=\partial_0 A_i-\partial_i A_0\,, \qquad  B_i=\frac{1}{2 }\epsilon_{i j k } \partial_j A_k\,,
\end{equation}
allows us to write the deformed Maxwell equations with an external source as
\begin{eqnarray}
&&\Omega^2(-\Box) \nabla\cdot\vec E=j_0\,,\quad  \Omega^2(-\Box) \vec \nabla\times \vec B= \Omega^2(-\Box)   \partial_0 \vec E + \vec j\,,\nonumber\\
 &&\Omega^2(-\Box) \vec \nabla\cdot\vec B=0\,,\quad \,\,\Omega^2(-\Box)\vec \nabla\times   \vec E=-\partial_0 \vec B\,.
  \label{eq:maxwell3}
\end{eqnarray}
Using the expressions above, the {Maxwell} equations for the fields are 
\begin{equation}
C(-\Box) \vec E=\nabla j_0+\partial_0 \vec j\,, \qquad C(-\Box)\vec B =-\vec\nabla\times \vec j\,.
\end{equation}The electromagnetic energy-tensor can be obtained from~\eqref{eq:action_EM2}, giving
\begin{equation}
    T_{\mu\nu}= \frac{1}{4}\eta_{\mu\nu} \tilde{F}_{\rho \sigma}\tilde{F}^{\rho \sigma}+\tilde{F}_{\mu \rho}{\tilde{F}^{\rho}}_ \nu\,.
\end{equation}

\begin{figure}[H]
    \centering
    \includegraphics[width=0.48\textwidth]{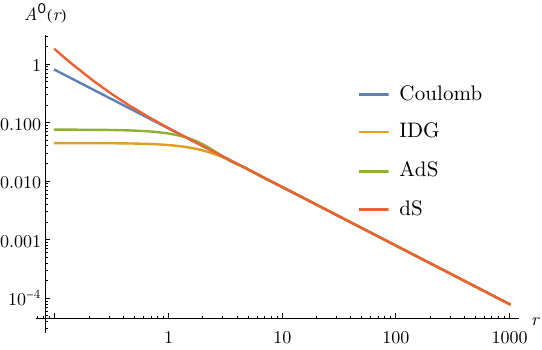}
    \caption{\small Electric scalar potential $A^0(r)$ as a function of the radius $r$ in the Coulomb case (blue), de Sitter (red), anti de Sitter (green) and IDG (yellow) for $q=\Lambda=1$. }
    \label{fig:A0rto0}
\end{figure}

We now consider the electric field of a point-like source. In such a case we deal with a source of the form $j^0(x)=q \, \delta^3(\vec{r}), \, j^i=0$, and the e.o.m. turn out to be exactly analytically solvable. The 0-component of the vector potential is the electric scalar potential, which can be written as
\begin{eqnarray}
    A^0(r) &=& -\frac{q}{2\pi^2} \int_0^\infty {\rm d} k\, \frac{ k^2}{C(-\vec{k}^2)}  \frac{\sin(kr)}{kr}.
\end{eqnarray}
{At this point, we need} to make a choice regarding the Casimir. When considering the metrics \eqref{eq:conformal_metric}, the dispersion relations become
\begin{align}
C_\text{dS}(p)\,=\,4 \Lambda^2 \arctanh^2\left(\frac{\sqrt{p^2}}{2 \Lambda}\right)\,=\,&m^2\,, \\ C_\text{AdS}(p)\,=\,4 \Lambda^2 \arctan^2\left(\frac{\sqrt{p^2}}{2 \Lambda}\right)\,=\,&m^2\,.
\label{eq:casimir_ds}
\end{align}
From here, it is easy to see that the electric scalar potential is constant for the AdS scenario and divergent for the dS one when $r\to 0$ (see FIG. \ref{fig:A0rto0}). 
{This seems to privilege AdS over dS momentum spaces since in a QG the divergences that appear in the usual QFT and GR should not be be present (for example, paradox information of black holes~\cite{Hawking:1976ra} could be alleviated if horizons and singularities are removed~\cite{Koshelev:2017bxd}). Therefore, our results are in agreement with~\cite{Matschull:1997du,Freidel:2003sp}, where it was shown that a QG in 2+1 dimensions should have the symmetries of an AdS momentum space.} 
If one chooses the Casimir operator derived from arguments of string theory used in most of the models of IDG, $C=e^{-\nabla^2/ \Lambda^2} \, \nabla^2$, the electric potential is given by
\begin{eqnarray}\label{Casimirs2}
    A^0_{IDG}(r)=  \frac{q}{2\pi^2} \int_0^\infty {\rm d}k\,  \frac{\sin(kr)}{kr\,  e^{k^2/\Lambda^2}}=\frac{q}{4\pi}Erf\left(\frac{r}{2}\right),
\end{eqnarray}
being $Erf$ the error function. We use this Casimir since it is widely used in the literature and shows a finite behavior of potentials at the origin of coordinates, so we can compare this result with the Casimirs proposed in this work (see~\cite{Relancio:2024loc} for further details). It can be checked that this model behaves at the origin like the AdS one.

Concerning the electric field, it can be obtained by means of $\vec{E}=-\vec{\nabla} A^0(r)$. An important conclusion that can be easily drawn is that the electric field does not diverge at $r \to 0$ for the AdS and the string  models but does for the dS one.

 \section{Nonlocality and soccer ball problem in DSR}
{The aim of this section is to discuss the nonlocality and soccer-ball problem appearing in DSR and how this is related to an infinite derivative QFT. In usual DSR theories, nonlocality comes from two different sources. Firstly, a non-commutative spacetime is associated to the deformed kinematics of DSR~\cite{KowalskiGlikman:2002jr,Battisti:2010sr}. This implies that the localizability of single (non-quantum) particles is lost since it is impossible to measure simultaneously the temporal and spatial coordinates of the particles. Then, there is a non-localizability of objects for different (translated) observers~\cite{Lizzi:2018qaf}. Secondly, a deformed conservation law for momentum forbids interactions to be always local for every observer (except the one located where the interaction takes place), since translations are also deformed. This is call the relative locality principle~\cite{AmelinoCamelia:2011bm}.  

An important problem of DSR theories is the so-called soccer-ball problem~\cite{Amelino-Camelia2011d}. It is based on the fact that, if there exists a deformed composition law of momenta, the total momentum of a system of several particles is different to the sum of their momenta. And this difference involve terms depending on the inverse of the high-energy scale. When considering low-energy particles, this correction is really small. However, when the system is formed by several particles, even if they are low energetic, this correction becomes non-negligible at all, being proportional to the quotient of the mass of the system and the high-energy scale.   Therefore, it seems to be a fundamental problem in DSR for macroscopic systems. This problem was circumvented in~\cite{Amelino-Camelia2011d} by considering  a redefinition of the high-energy scale as a function of the number of particles $N$, i.e., the authors porposed the replacement $\Lambda\to N\Lambda$ when a macroscopic system is regarded (see~\cite{Carmona:2023luz} for a different proposal of resolution). 

Our proposal of an infinite derivative QFT is based on the field theory on curved momentum spaces developed in~\cite{Franchino-Vinas:2022fkh}. As it was shown in~\cite{Carmona:2019fwf}, there is a clear connection between a curved momentum space and a deformed relativistic kinematics (as discussed in the introduction). Moreover, there are several papers connecting a non-commutative spacetime  with a curved momentum space,  and then to a deformed relativistic kinematics~\cite{Arzano:2010kz,Carmona:2019fwf,Relancio:2021ahm,Wagner:2021bqz}. Hence, the non-locality described in our QFT framework should be identified with the effects of a noncommutative spacetime~\cite{Buoninfante2019}, since we are not dealing with interactions. Therefore, we present here a field theory in which the nonlocal effects of the noncommutativity of space-time coordinates is replaced by the nonlocality given by the action of infinite derivatives over the fields.

Furthermore, nonlocal effects due to a deformed composition law are expected to appear when introducing interactions, but they should be connected to the nonlocality associated to the relative locality principle. Interestingly, in~\cite{Buoninfante:2018gce} it is discussed that, when considering interactions, the high-energy scale involved in the process becomes $N$-dependent. This is exactly the solution proposed in~\cite{Amelino-Camelia2011d} to solve the soccer-ball problem. It is worth mentioning that in our work, we start with the free part of the theory (and we also consider the presence of electromagnetic sources) to set the building blocks of a complete infinite derivative QFT. In such a complete QFT,  interactions should be described by a deformed composition law of momenta (compatible with a deformed dispersion relation). This theory should describe the nonlocality of both single- and multi-particle systems, alleviating the soccer-ball problem at the same time. We hope to explore this in future works.

}

   \section{Conclusions and outlook}
\label{sec:conclusions}
In this {work}, we propose a novel way to consider QFT in DSR scenarios. This construction, based on the geometrical arguments of a curved momentum space, leads to a nonlocal QFT. {A significant advantage of our proposal is that it integrates other quantum gravity scenarios into the same framework,} such as string and causal set theories, and, therefore, one can understand the deformation of QFT in such theories as a deformation of the SR kinematics described by a curved momentum space. This interpretation has not been made in the literature so far. Furthermore, the developments in our framework can be extrapolated {for other proposals of IDG}. 

We have described the Klein--Gordon, Dirac, and electromagnetic Lagrangians in our approach. We have also found the deformed Maxwell equations in presence of an external source. Another strength of our construction is that it can be easily derived from the usual relativistic QFT by replacing the usual derivatives $\partial_\mu$ by an infinite function of them $\tilde \partial_\mu=\partial_\mu \Omega(\Box).$ 

Moreover, our scheme restricts the possible kinematical models in DSR scenarios, leaving only space to Snyder kinematics, which has at {its} base a Lorentz invariance for a single- and multi-particle systems.  For a maximally symmetric and conformally Minkowski momentum metric, which has been considered in the literature to be advantaged from a geometrical point of view, we find that the  electric scalar potential does not diverge at the origin for the AdS model. This is in agreement with previous results establishing a relationship between a 2+1 QG theory and the symmetries of an AdS momentum space. It is important to note that there are infinite possible choices of coordinates in a AdS space. One possible criteria to restrict the possible models (and then the bases of the kinematics) is to impose the absence of ghosts in the theory. This discussion is beyond the scope of this paper and is left for further research.

Finally, we expect that the results of this {work} paves the way to a full description of all single (non-interaction) effects of QFT in DSR scenarios. The main pillar of these theories is a deformed conservation for momenta which will be a milestone we hope to explore in future works {regarding interaction theory}. Moreover, the exploration of the gravitational Lagrangian and the extension of this work to curved spacetimes is left for further research.

\paragraph{Acknowledgments.}
{This work has been supported by the grant PID2023-148373NB-I00 funded by MCIN /AEI /10.13039/501100011033 / FEDER, UE, and by the Q-CAYLE Project funded by the Regional Government of Castilla y Le\'on (Junta de Castilla y Le\'on) and by the Ministry of Science and Innovation (MCIN) through the European Union funds NextGenerationEU (PRTR C17.I1). The authors thank J. L. Cortés and S. Liberati for fruitful discussions.}

%


\end{document}